\documentclass[12pt]{article}

\usepackage{amsmath,amssymb}
\usepackage{amsmath,amscd}
\usepackage{bm}
\usepackage{mathrsfs}
\usepackage{latexsym}
\usepackage{multirow}
\usepackage{float}
\usepackage[usenames]{color}
\usepackage{indentfirst}  
\usepackage{cite}

\allowdisplaybreaks[4]

\newcommand{{\Slashp}}{p\!\!\!\!\!\big/}
\newcommand{{\Slashq}}{q\!\!\!\!\!\big/}

\newcommand{\LBox}{\mbox{\Large$\Box$}}

\makeatletter
\@addtoreset{equation}{section}

\makeatother

\begin{document}

\begin{titlepage}
\thispagestyle{empty}
~~\\
\vspace{5mm} 

\begin{center}
{\LARGE Conjugate boundary condition, hidden particles,\\
and gauge-Higgs inflation}
\end{center}

\begin{center}
\lineskip .45em
\vskip1.5cm
{\large Yugo Abe$^a$\footnote{E-mail: yugo@riko.shimane-u.ac.jp}, 
Yuhei Goto$^b$\footnote{E-mail:  14st302a@shinshu-u.ac.jp}, 
Yoshiharu Kawamura$^b$\footnote{E-mail: haru@azusa.shinshu-u.ac.jp},\\
and Yasunari Nishikawa$^b$\footnote{E-mail: 15sm210a@shinshu-u.ac.jp}}

\vskip 1.5em
${}^a\,${\large\itshape Graduate School of Science and Engineering, 
Shimane University, Matsue 690-8504, Japan}\\[1mm]
${}^a\,${\large\itshape Department of Physics, Tokyo Institute of Technology,
Tokyo 152-8551, Japan}\\[1mm]
${}^b\,${\large\itshape Department of Physics, Shinshu University, Matsumoto 390-8621, Japan}

 \vskip 4.5em
\end{center}



\begin{abstract}
We propose an idea that 
hidden particles can be separated
according to gauge quantum numbers
from the visible ones by the difference of boundary conditions
on extra dimensions.
We formulate 5-dimensional gauge theories 
yielding conjugate boundary conditions besides ordinary ones
on $S^1/Z_2$, and examine physical implications 
concerning hidden particles
on an extension of the standard model 
coexisting different types of boundary conditions.
A model with conjugate boundary conditions is applied
on a gauge-Higgs inflation scenario.
\end{abstract}

\end{titlepage}

\section{Introduction}

There are several riddles that cannot be solved
in the framework of the standard model (SM),
but any evidences from new physics such as supersymmetry, 
compositeness and extra dimensions have not been discovered.
Here, the SM means the extended model including
the modifications with massive neutrinos.

One of big issues in physics beyond the SM
is to disclose the identity of unknown
particles such as dark matter and inflaton.
Because it is hard to detect such hidden particles directly,
they are supposed to interact with the SM particles weakly.
A typical example is a scenario based on 
the Weakly Interacting Massive Particle (WIMP) assumption for dark matter,
and possible dark matter candidates are the lightest superparticle (LSP)
in a supersymmetric extension of the SM
and the lightest Kaluza-Klein mode in an extra-dimensional one.

In this paper, we consider an extreme case 
that hidden particles are singlets of the SM gauge symmetries 
and the SM particles are singlets of gauge symmetries
in a hidden sector.
Here, the hidden sector stands for a sector that
contains singlets of the SM gauge group besides graviton.
In this case, we have a question $\lq\lq$how is such a subtle
separation of gauge quantum numbers realized naturally?''
We present an idea that 
{\it hidden particles can be separated
according to gauge quantum numbers
from the visible ones 
by the difference of boundary conditions (BCs) 
on extra dimensions}, as a possible answer.
To illustrate our idea, we construct 5-dimensional (5D) 
gauge theories yielding conjugate BCs besides ordinary ones on $S^1/Z_2$,
and show that the separation of visible and hidden particles
can be realized in gauge interactions
using a 5D extension of the SM with an extra $U(1)$ gauge symmetry
coexisting different types of BCs.
Furthermore, we study models that 
hidden particles relating to conjugate BCs 
are identified with dark matter or inflaton. 

This paper is organized as follows.
In the next section, we formulate a 5D $U(1)$ gauge theory 
yielding conjugate BCs on $S^1/Z_2$.
We examine physical implications on a theory 
coexisting different types of BCs in Sect.~3, 
and apply a model with conjugate BCs 
on a gauge-Higgs inflation scenario in Sect.~4.
In the last section, we give conclusions and discussions.

\section{Conjugate boundary condition on $S^1/Z_2$}
\label{CBC}

The space-time is assumed to be factorized into 
a product of 4-dimensional (4D) Minkowski space-time $M^4$ 
and the orbifold $S^1/Z_2$, whose coordinates are denoted by 
$x^\mu$ (or $x$) $(\mu=0,1,2,3)$ and $y~(=x^5)$, respectively.
The 5D notation $x^M$ $(M=0,1,2,3,5)$ is also used.
The $S^1/Z_2$ is obtained 
by dividing the circle $S^1$ (with the identification $y\sim y+2\pi R$)
by the $Z_2$ transformation $y \rightarrow -y$.
Then, the point $y$ is identified with $-y$ on $S^1/Z_2$,
and the space is regarded as an interval with length $\pi R$
($R$ being the radius of $S^1$).

In the following, we formulate a 5D $U(1)$ gauge theory
with ordinary BCs and that with conjugate BCs, respectively.

\subsection{Ordinary boundary condition}
\label{OBC}

For completeness, we explain ordinary BCs
using the 5D $U(1)$ gauge theory
whose Lagrangian density is given by
\begin{eqnarray}
\mathscr{L}_B= (D_M \varphi)^{*} (D^M \varphi) - m_{\varphi}^2 |\varphi|^2
 + \bar{\psi}(i\Gamma^MD_M-m_{\psi})\psi
 - \frac{1}{4} B_{MN}B^{MN},
\label{LB}
\end{eqnarray}
where $D_M=\partial_M-ig_5q_{\varphi} B_M$ 
for a complex scalar field $\varphi = \varphi(x, y)$,
$*$ stands for the complex conjugation,
$D_M=\partial_M-ig_5q_{\psi} B_M$ 
for a 5D spinor field $\psi = \psi(x, y)$,
and $m_{\varphi}$ and $m_{\psi}$ are 
masses of $\varphi$ and $\psi$, respectively.
The $\bar{\psi}$ is the Dirac conjugate of $\psi$
defined by $\bar{\psi} \equiv \psi^\dagger \gamma^0$.
The $g_5$ is a 5D gauge coupling constant,
$B_M(=B_M(x,y))$ is a 5D $U(1)$ gauge boson,
$B_{MN}$ is the gauge strength 
defined by $B_{MN} = \partial_M B_N - \partial_N B_M$,
and $q_{\varphi}$ and $q_{\psi}$ are 
$U(1)$ charges of $\varphi$ and $\psi$, respectively.
The 5D gamma matrices $\Gamma^M$ are given by 
\begin{eqnarray}
\Gamma^{\mu}=\gamma^{\mu},~~ \Gamma^5=i\gamma^5,
\label{Gamma}
\end{eqnarray}
using 4D gamma matrices $\gamma^\mu$ and 
$\gamma^5 = i \gamma^0 \gamma^1 \gamma^2 \gamma^3$,
and they satisfy the algebraic relation 
$\{\Gamma^M, \Gamma^N\}=2\eta^{MN}$
with $\eta^{MN} = {\rm diag}(1,-1,-1,-1,-1)$.

From the requirement that 
the Lagrangian density should be invariant under the translation
$y \to y + 2\pi R$ and
the $Z_2$ transformation $y \to -y$
or it should be a single-valued function on the 5D space-time,
the BCs of fields on $S^1/Z_2$ are determined up to some parameters
such as intrinsic $Z_2$ parities.

Because the derivative $\partial_M$ is invariant under $y \to y + 2\pi R$,
$D_M$ should also be invariant.
Under $y \to -y$,
$\partial_M$ transforms as $\partial_\mu \to \partial_\mu$
and $\partial_5 \to - \partial_5$, and hence
$D_M$ is ordinarily supposed to transform as
\begin{eqnarray}
D_{\mu} \to D_{\mu},~~ D_5 \to - D_5.
\label{DM-OBC}
\end{eqnarray}
Then, the BCs of $B_M$ are determined as
\begin{eqnarray}
&~& B_M(x, y+2\pi R)=B_M(x, y),
\label{BM-OBC-T}\\
&~& B_\mu(x, -y)=B_\mu(x, y),~~
B_5(x, -y)=-B_5(x, y),
\label{BM-OBC-Z2}
\end{eqnarray}
and $B_M$ are given by the Fourier expansions:
\begin{eqnarray}
&~& B_\mu(x, y)=\frac{1}{\sqrt{2\pi{R}}}B_{\mu}^{(0)}(x)
+\frac{1}{\sqrt{\pi{R}}}\sum_{n=1}^{\infty}B_{\mu}^{(n)}(x)\cos\frac{ny}{R},
\label{Bmu-exp}\\
&~& B_5(x, y)=\frac{1}{\sqrt{\pi{R}}}\sum_{n=1}^{\infty}
B_{5}^{(n)}(x)\sin\frac{ny}{R}.
\label{B5-exp}
\end{eqnarray}
Note that a zero mode (or a $y$-independent part) of $B_5$ is absent.

The BCs of $\varphi$ and $\psi$ are determined as
\begin{eqnarray}
&~& \varphi(x, y+2\pi R)=e^{i \beta_{\varphi}} \varphi(x, y),~~
\varphi(x, -y)=\eta_{\varphi} \varphi(x, y),
\label{varphi-OBC}\\
&~& \psi(x, y+2\pi R)=e^{i \beta_{\psi}} \psi(x, y),~~
\psi(x, -y)=\eta_{\psi} i \Gamma^5 \psi(x, y),
\label{psi-OBC}
\end{eqnarray}
where $\beta_{\varphi}$ and $\beta_{\psi}$ take $0$ or $\pi$.
The $\eta_{\varphi}$ and $\eta_{\psi}$ are the intrinsic $Z_2$ parity
of $\varphi$ and $\psi$, respectively, and they take $1$ or $-1$.
From (\ref{varphi-OBC}), $\varphi$ is given by the Fourier expansions:
\begin{eqnarray}
&~& \varphi(x, y)=\frac{1}{\sqrt{2\pi{R}}}\varphi^{(0)}(x)
+\frac{1}{\sqrt{\pi{R}}}\sum_{n=1}^{\infty}\varphi^{(n)}(x)\cos\frac{ny}{R},
\label{varphi-exp1}\\
&~& \varphi(x, y)=\frac{1}{\sqrt{\pi{R}}}\sum_{n=1}^{\infty}
\varphi^{(n)}(x)\sin\frac{ny}{R},
\label{varphi-exp2}\\
&~& \varphi(x, y)=\frac{1}{\sqrt{\pi{R}}}\sum_{n=1}^{\infty}
\varphi^{(n)}(x)\cos\frac{\left(n-\frac{1}{2}\right)y}{R},
\label{varphi-exp3}\\
&~& \varphi(x, y)=\frac{1}{\sqrt{\pi{R}}}\sum_{n=1}^{\infty}
\varphi^{(n)}(x)\sin\frac{\left(n-\frac{1}{2}\right)y}{R}
\label{varphi-exp4}
\end{eqnarray}
for $(\beta_{\varphi}, \eta_{\varphi})$ is $(0, 1)$, $(0, -1)$,
$(\pi, 1)$ and $(\pi, -1)$, respectively.
In a similar way, from (\ref{psi-OBC}), 
$\psi$ is also given as Fourier expansions.
Here, we give only a case with $m_{\psi} = 0$,
$\beta_{\psi}=0$ and $\eta_{\psi}=1$,
\begin{eqnarray}
&~& \psi_{\rm L}(x, y)=\frac{1}{\sqrt{2\pi{R}}}\psi_{\rm L}^{(0)}(x)
+\frac{1}{\sqrt{\pi{R}}}\sum_{n=1}^{\infty}\psi_{\rm L}^{(n)}(x)\cos\frac{ny}{R},
\label{phi-exp1}\\
&~& \psi_{\rm R}(x, y)=\frac{1}{\sqrt{\pi{R}}}\sum_{n=1}^{\infty}
\psi_{\rm R}^{(n)}(x)\sin\frac{ny}{R},
\label{psi-exp2}
\end{eqnarray}
where $\psi_{\rm L}$ and $\psi_{\rm R}$ are 
2-component spinor fields whose 4D chirality 
(the eigenvalue of $\gamma^5$)
is $-1$ and $1$, respectively.

\subsection{Conjugate boundary condition}
\label{subCBC}

Let us study another BCs using the 5D $U(1)$ gauge theory
whose Lagrangian density is given by
\begin{eqnarray}
\mathscr{L}_{C} = (D_M \tilde{\varphi})^{*} (D^M \tilde{\varphi}) 
 - m_{\tilde{\varphi}}^2 |\tilde{\varphi}|^2
 + \bar{\tilde{\psi}}(i\Gamma^MD_M-m_{\tilde{\psi}})\tilde{\psi}
 - \frac{1}{4} C_{MN}C^{MN},
\label{LC}
\end{eqnarray}
where $D_M=\partial_M-i\tilde{g}_5 \tilde{q}_{\tilde{\varphi}} C_M$ 
for a complex scalar field $\tilde{\varphi} = \tilde{\varphi}(x, y)$,
$D_M=\partial_M-i\tilde{g}_5 \tilde{q}_{\tilde{\psi}} C_M$ 
for a 5D spinor field $\tilde{\psi} = \tilde{\psi}(x, y)$,
and $m_{\tilde{\varphi}}$ and $m_{\tilde{\psi}}$ are 
masses of $\tilde{\varphi}$ and $\tilde{\psi}$, respectively.
The $\tilde{g}_5$ is a 5D gauge coupling constant,
$C_M$ is a 5D $U(1)$ gauge boson,
$C_{MN}$ is the gauge strength 
defined by $C_{MN} = \partial_M C_N - \partial_N C_M$,
and $\tilde{q}_{\tilde{\varphi}}$ and $\tilde{q}_{\tilde{\psi}}$ are 
$U(1)$ charges of $\tilde{\varphi}$ and $\tilde{\psi}$, respectively.

Under $y \to y + 2\pi R$, let $D_M$ be invariant.
Under $y \to -y$, let $D_M$ transform as
\begin{eqnarray}
D_{\mu} \to D_{\mu}^*,~~ D_5 \to - D_5^*.
\label{DM-CBC}
\end{eqnarray}
Then, the BCs of $C_M$ are determined as
\begin{eqnarray}
&~& C_M(x, y+2\pi R)=C_M(x, y),
\label{CM-CBC-T}\\
&~& C_{\mu}(x, -y)=-C_\mu(x, y),~~
C_5(x, -y)=C_5(x, y),
\label{CM-CBC-Z2}
\end{eqnarray}
and $C_M$ are given by the Fourier expansions:
\begin{eqnarray}
&~& C_\mu(x, y)=\frac{1}{\sqrt{\pi{R}}}\sum_{n=1}^{\infty}
C_{\mu}^{(n)}(x)\sin\frac{ny}{R},
\label{Cmu-exp}\\
&~& C_5(x, y)=\frac{1}{\sqrt{2\pi{R}}}C_{5}^{(0)}(x)
+\frac{1}{\sqrt{\pi{R}}}\sum_{n=1}^{\infty}C_{5}^{(n)}(x)\cos\frac{ny}{R}.
\label{C5-exp}
\end{eqnarray}
Note that $C_{5}$ has even $Z_2$ parities, 
and its zero mode $C_5^{(0)}$ survives and becomes a dynamical field.

The BCs of $\tilde{\varphi}$ and $\tilde{\psi}$ 
are determined as
\begin{eqnarray}
&~& \tilde{\varphi}(x, y+2\pi R)
=e^{i \beta_{\tilde{\varphi}}} \tilde{\varphi}(x, y),~~
\tilde{\varphi}(x, -y)= 
\tilde{\varphi}^*(x, y),
\label{varphi-CBC}\\
&~& \tilde{\psi}(x, y+2\pi R)
=e^{i \beta_{\tilde{\psi}}} \tilde{\psi}(x, y),~~
\tilde{\psi}(x, -y)= i \tilde{\psi}^{c}(x, y),
\label{psi-CBC}
\end{eqnarray}
where $\beta_{\tilde{\varphi}}$ and $\beta_{\tilde{\psi}}$ 
are arbitrary real constants and
$\tilde{\psi}^c = e^{i\gamma_{\rm c}} \Gamma^2 \tilde{\psi}^*$.
The $\tilde{\psi}^c$ corresponds to
a charge conjugation of $\tilde{\psi}$ on 4D space-time,
and $\gamma_{\rm c}$ is an arbitrary real number.
A set of BCs (\ref{CM-CBC-T}), (\ref{CM-CBC-Z2}),
(\ref{varphi-CBC}) and (\ref{psi-CBC}) 
corresponds to that in a $U(1)$ case of the orbifold breaking 
by outer automorphisms~\cite{HM},
and we refer to
such BCs relating particles with a representation $\bf{R}$
to that with the conjugated one $\overline{\bf{R}}$ 
as {\it conjugate BCs}~\cite{HKO}.

We find that the complex scalar field part of $\mathscr{L}_C$
is single-valued from the translation 
and the $Z_2$ transformation properties such that
\begin{eqnarray}
\hspace{-2cm}&~&
 (\partial_M-i\tilde{g}_5 \tilde{q}_{\tilde{\varphi}} C_M(x, y))\tilde{\varphi}(x, y)
\to 
(\partial_M-i\tilde{g}_5 \tilde{q}_{\tilde{\varphi}} C_M(x, y+2\pi R))
\tilde{\varphi}(x, y+2\pi R)
\nonumber \\
\hspace{-2cm}&~& ~~~~~~~~~~~~~~~~~~~~~~~~~~~~~~~~~~~~~~~~~~
 = e^{i \beta_{\tilde{\varphi}}} 
(\partial_M-i\tilde{g}_5 \tilde{q}_{\tilde{\varphi}} C_M(x, y))\tilde{\varphi}(x, y)
\label{DMvarphi-CBC-T}
\end{eqnarray}
and
\begin{eqnarray}
\hspace{-1.5cm}&~& 
(\partial_{\mu}-i\tilde{g}_5 \tilde{q}_{\tilde{\varphi}} C_{\mu}(x, y))
\tilde{\varphi}(x, y) 
\to 
(\partial_{\mu}-i\tilde{g}_5 \tilde{q}_{\tilde{\varphi}} C_{\mu}(x, -y))
\tilde{\varphi}(x, -y)
\nonumber \\
\hspace{-1.5cm}&~& ~~~~~~~~~~~~~~~~~~~~~~~~~~~~~~~~~~~~~~~~ 
= (\partial_{\mu}+i\tilde{g}_5 \tilde{q}_{\tilde{\varphi}} C_{\mu}(x, y))
\tilde{\varphi}^*(x, y)
\nonumber \\
\hspace{-1.5cm}&~& ~~~~~~~~~~~~~~~~~~~~~~~~~~~~~~~~~~~~~~~~  
= (D_{\mu}\tilde{\varphi}(x, y))^*,
\label{Dmuvarphi-CBC-Z2}\\
\hspace{-1.5cm}&~& 
(\partial_{5}-i\tilde{g}_5 \tilde{q}_{\tilde{\varphi}} C_{5}(x, y))
\tilde{\varphi}(x, y) 
\to 
(-\partial_{5}-i\tilde{g}_5 \tilde{q}_{\tilde{\varphi}} C_{5}(x, -y))
\tilde{\varphi}(x, -y)
\nonumber \\
\hspace{-1.5cm}&~& ~~~~~~~~~~~~~~~~~~~~~~~~~~~~~~~~~~~~~~~~ 
= -(\partial_{5}+i\tilde{g}_5 \tilde{q}_{\tilde{\varphi}} C_{5}(x, y))
\tilde{\varphi}^*(x, y)
\nonumber \\
\hspace{-1.5cm}&~& ~~~~~~~~~~~~~~~~~~~~~~~~~~~~~~~~~~~~~~~~ 
= -(D_{5}\tilde{\varphi}(x, y))^*,
\label{D5varphi-CBC-Z2}
\end{eqnarray}
respectively.
From (\ref{varphi-CBC}), $\tilde{\varphi}$ 
is given by the Fourier expansion:
\begin{eqnarray}
\tilde{\varphi}(x, y)=
\frac{1}{2\sqrt{\pi{R}}}\sum_{n=-\infty}^{\infty}
\tilde{\varphi}^{(n)}(x) e^{i\frac{2\pi n + \beta_{\tilde{\varphi}}}{2\pi R}y},
\label{tildevarphi-exp}
\end{eqnarray}
where $\tilde{\varphi}^{(n)}(x)$ are 4D real scalar fields
($\tilde{\varphi}^{(n)*}(x) = \tilde{\varphi}^{(n)}(x)$).

In a similar way,
we find that the spinor field part of $\mathscr{L}_C$
is also single-valued from the transformation properties such that
\begin{eqnarray}
&~& \bar{\tilde{\psi}}(x, y)
\left\{i\Gamma^M(\partial_M-i\tilde{g}_5 \tilde{q}_{\tilde{\psi}} C_M(x,y))
-m_{\tilde{\psi}}\right\}\tilde{\psi}(x, y)
\nonumber \\
&~& ~~ \to \bar{\tilde{\psi}}(x, y+2\pi R)
\left\{i\Gamma^M(\partial_M-i\tilde{g}_5 \tilde{q}_{\tilde{\psi}} C_M(x,y+2\pi R))
-m_{\tilde{\psi}}\right\}\tilde{\psi}(x, y+2\pi R)
\nonumber \\
&~& ~~~~~~ 
= \bar{\tilde{\psi}}(x, y)
\left\{i\Gamma^M(\partial_M-i\tilde{g}_5 \tilde{q}_{\tilde{\psi}} C_M(x,y))
-m_{\tilde{\psi}}\right\}\tilde{\psi}(x, y)
\label{DMpsi-CBC-T}
\end{eqnarray}
and
\begin{eqnarray}
&~& \bar{\tilde{\psi}}(x, y)
\left\{i\Gamma^M(\partial_M-i\tilde{g}_5 \tilde{q}_{\tilde{\psi}} C_M(x,y))
-m_{\tilde{\psi}}\right\}\tilde{\psi}(x, y)
\nonumber \\
&~& ~~ \to 
\bar{\tilde{\psi}}(x, -y)\left\{i\Gamma^M(\partial_M-i\tilde{g}_5 \tilde{q}_{\tilde{\psi}} C_M(x,-y))
-m_{\tilde{\psi}}\right\}
\tilde{\psi}(x, -y)
\nonumber \\
&~& ~~~~~~ = \bar{\tilde{\psi}}^c(x, y)
\left\{i\Gamma^{\mu} 
(\partial_{\mu}+i\tilde{g}_5 \tilde{q}_{\tilde{\psi}} C_{\mu}(x,y)) \right.
\nonumber \\
&~& ~~~~~~~~~~~~~~~~~~~~~~
\left. - i\Gamma^{5}
(\partial_{5}+i\tilde{g}_5 \tilde{q}_{\tilde{\psi}} C_{5}(x,y))
-m_{\tilde{\psi}}\right\}\tilde{\psi}^c(x, y)
\nonumber \\
&~& ~~~~~~ 
= \bar{\tilde{\psi}}(x, y)
\left\{i\Gamma^M(\partial_M-i\tilde{g}_5 \tilde{q}_{\tilde{\psi}} C_M(x,y))
-m_{\tilde{\psi}}\right\}\tilde{\psi}(x, y),
\label{DMpsi-CBC-Z2}
\end{eqnarray}
where we use the following relations concerning
4-component spinor fields $\tilde{\psi}$:
\begin{eqnarray}
&~& \bar{\tilde{\psi}}^c \gamma^{\mu} \partial_{\mu} \tilde{\psi}^c
= \bar{\tilde{\psi}} \gamma^{\mu} \partial_{\mu} \tilde{\psi},~~
\bar{\tilde{\psi}}^c \gamma^{5} \partial_{5} \tilde{\psi}^c
= - \bar{\tilde{\psi}} \gamma^{5} \partial_{5} \tilde{\psi},
\nonumber \\
&~& \bar{\tilde{\psi}}^c \gamma^{\mu} \tilde{\psi}^c
= - \bar{\tilde{\psi}} \gamma^{\mu} \tilde{\psi},~~
\bar{\tilde{\psi}}^c \gamma^{5} \tilde{\psi}^c
= \bar{\tilde{\psi}} \gamma^{5} \tilde{\psi},~~
\bar{\tilde{\psi}}^c \tilde{\psi}^c
= \bar{\tilde{\psi}} \tilde{\psi}.
\label{cc-relations}
\end{eqnarray}
From (\ref{psi-CBC}), $\tilde{\psi}$ is given by the Fourier expansion:
\begin{eqnarray}
\tilde{\psi}(x, y)=
\frac{1}{2\sqrt{\pi{R}}}\sum_{n=-\infty}^{\infty}
\left(
\begin{array}{c}
\tilde{\xi}_{\alpha}^{(n)}(x)  \\
i \bar{\tilde{\xi}}^{(n)\dot{\alpha}}(x) 
\end{array}
\right)
e^{i\frac{2\pi n + \beta_{\tilde{\psi}}}{2\pi R}y},
\label{tildepsi-exp}
\end{eqnarray}
where $\tilde{\xi}_{\alpha}^{(n)}(x)$ are 4D 2-component spinor fields,
and $\alpha$ and $\dot{\alpha}$ are spinor indices.
Hereafter, we omit the spinor indices.

By inserting the mode expansions (\ref{tildevarphi-exp}) and (\ref{tildepsi-exp})
into the 5D action $S_{\rm 5D} = \int \mathscr{L}_C d^5x$, 
we obtain the following 4D action: 
\begin{eqnarray}
&~& S_{\rm 4D}
=\int{d^4x}
\left[- \sum_{n=-\infty}^{\infty} \frac{1}{2} \tilde{\varphi}^{(n)}
\left\{\raisebox{-0.9mm}{\LBox}+\left(\frac{2\pi n
+ \beta_{\tilde{\varphi}} - q_{\tilde{\varphi}} \theta}
{2\pi R}\right)^2+m_{\tilde{\varphi}}^{2}\right\}\tilde{\varphi}^{(n)}\right.
\nonumber \\
&~& ~~~~~~~~~~~~~~~~~~~ + 
\sum_{n=-\infty}^{\infty} \frac{1}{2} 
\biggl\{\tilde{\xi}^{(n)} i\sigma^{\mu} \partial_{\mu} \bar{\tilde{\xi}}^{(n)} 
+ \bar{\tilde{\xi}}^{(n)} i\overline{\sigma}^{\mu} \partial_{\mu} \tilde{\xi}^{(n)}
\nonumber \\
&~& ~~~~~~~~~~~~~~~~~~~ +
\left(\frac{2\pi n+\beta_{\tilde{\psi}} - \tilde{q}_{\tilde{\psi}} \theta}{2\pi R} 
+ i m_{\tilde{\psi}}\right) \tilde{\xi}^{(n)} \tilde{\xi}^{(n)}
\nonumber \\
&~& ~~~~~~~~~~~~~~~~~~~ +
\left.
\left(\frac{2\pi n+\beta_{\tilde{\psi}} - \tilde{q}_{\tilde{\psi}} \theta}{2\pi R} 
- i m_{\tilde{\psi}}\right) \bar{\tilde{\xi}}^{(n)} \bar{\tilde{\xi}}^{(n)}
\right\}
\Bigg] + \cdots,
\label{S-4D}
\end{eqnarray}
where $\sigma^{\mu} = (I, \bm{\sigma})$,
$\overline{\sigma}^{\mu} = (I, -\bm{\sigma})$ 
($\bm{\sigma} = (\sigma_x, \sigma_y, \sigma_z)$
are Pauli matrices),
$\theta$ is the Wilson line phase
defined by 
\begin{eqnarray}
\theta =  \tilde{g}_5 \int_{-\pi R}^{\pi R} \frac{1}{\sqrt{2\pi R}}C_5^{(0)} dy 
= \sqrt{2\pi R} \tilde{g}_5 C_5^{(0)},
\label{theta}
\end{eqnarray}
and the ellipsis stands for 
parts containing Kaluza-Klein modes of gauge bosons
and the kinetic term of $C_5^{(0)}$.
Note that the $U(1)$ gauge symmetry is broken by orbifolding,
and $\theta$ is a remnant of the $U(1)$.

\section{Why hidden}
\label{WH}

In order to obtain some hints
to explore the origin of dark matter and the identity of inflaton
and to address the reason for their existence,
we search for an factor that it is hard to detect hidden particles
based on the following assumptions.
\begin{itemize}
\item There is an extra gauge group $G_{\rm hidden}$
other than the SM one $G_{\rm SM}$
(or some extension such as a grand unified group $G_{\rm GUT}$), 
and $G_{\rm hidden}$ leaves little trace behind
around the terascale.

\item Hidden particles such as dark matter and inflaton 
possess gauge quantum numbers of $G_{\rm hidden}$
or are some components of gauge bosons in a hidden sector,
and they are gauge singlets of $G_{\rm SM}$ (or $G_{\rm GUT}$).

\item The SM particles are gauge singlets of $G_{\rm hidden}$.
\end{itemize}

Gauge quantum numbers are suitably assigned 
to construct a realistic model, but in most cases,
it would be done without any foundation
except for symmetry principle.
We expect a reason or a mechanism that
a subtle separation of gauge quantum numbers 
in the above assumptions
is realized naturally, and propose a hypothesis that 
{\it hidden particles can be separated
according to gauge quantum numbers
from the visible ones by the difference of BCs
on extra dimensions}.\footnote{
According to a similar idea that a dark matter possesses
different features from the SM particles on extra dimensions,
a truncated-inert-doublet model has been constructed
that the SM ones belong to $Z_2$ even zero modes
and the dark matter is one of $Z_2$ odd zero modes
on a warped extra dimension~\cite{AGGJ}.
}

To embody our hypothesis, we consider a 5D theory with 
$G_{\rm SM} \times U(1)_C$ gauge group as 
an extension of the SM with an extra $U(1)$ gauge boson $C_M=C_M(x, y)$
and an extra matter $\tilde{\varphi}=\tilde{\varphi}(x, y)$.
For simplicity, we pay attention to scalar fields
and $U(1)$ gauge bosons and treat the Lagrangian density,
\begin{eqnarray}
&~& \mathscr{L}_{\rm 5D}
= (D_M H)^{*} (D^M H) - m_{H}^2 |H|^2
 - \frac{1}{4} B_{MN}B^{MN}
\nonumber \\
&~& ~~~~~~ + (D_M \tilde{\varphi})^{*} (D^M \tilde{\varphi}) 
 - m_{\tilde{\varphi}}^2 |\tilde{\varphi}|^2
 - \frac{1}{4} C_{MN}C^{MN}
\nonumber \\
&~& ~~~~~~ - \lambda \left(|H|^2\right)^2
- \lambda_{\tilde{\varphi}} \left(|\tilde{\varphi}|^2\right)^2
- \lambda_{\rm mix} |H|^2 |\tilde{\varphi}|^2
+ \cdots,
\label{L5D}
\end{eqnarray}
where $H=H(x, y)$ is 5D complex scalar field
containing the SM Higgs doublet as its zero mode ($H^{(0)}$), and
$\lambda$, $\lambda_{\tilde{\varphi}}$
and $\lambda_{\rm mix}$ are quartic couplings of scalar fields.

If $B_{M}$ (the 5D extension of the $U(1)_Y$ gauge boson
in the SM) satisfies the BCs 
such as (\ref{BM-OBC-T}) and (\ref{BM-OBC-Z2})
and $C_{M}$ satisfies
the BCs such as (\ref{CM-CBC-T}) and (\ref{CM-CBC-Z2}),
$H$ and $\tilde{\varphi}$ cannot own both non-zero $U(1)$ charges.
In other words, $H$ is separated from $\tilde{\varphi}$
in gauge interactions through the difference of BCs.

After the dimensional reduction,
we obtain the following 4D Lagrangian density for zero modes
$H^{(0)}$, $\tilde{\varphi}^{(0)}$, $B_{\mu}^{(0)}$ and $C_5^{(0)}$,
at the tree level,
\begin{eqnarray}
&~& \mathscr{L}_{\rm 4D}^{(0)}
= (D_{\mu}^{(0)} H^{(0)})^{*} (D^{(0)\mu} H^{(0)}) 
- m_{H}^2 |H^{(0)}|^2
 - \frac{1}{4} B_{\mu\nu}^{(0)}B^{(0)\mu\nu}
\nonumber \\
&~& ~~~~~~ + \frac{1}{2} \partial_{\mu} \tilde{\varphi}^{(0)}
\partial^{\mu} \tilde{\varphi}^{(0)} 
 - \frac{1}{2} \left\{m_{\tilde{\varphi}}^2 
+ \left(\frac{\beta_{\tilde{\varphi}} 
- \tilde{q}_{\tilde{\varphi}} \theta}{2\pi R}\right)^2\right\}
 (\tilde{\varphi}^{(0)})^2
 + \frac{1}{2} \partial_{\mu} C_5^{(0)} \partial^{\mu} C_5^{(0)}
\nonumber \\
&~& ~~~~~~ - \lambda \left(|H^{(0)}|^2\right)^2
- \frac{1}{4} \lambda_{\tilde{\varphi}} (\tilde{\varphi}^{(0)})^4
- \frac{1}{2} \lambda_{\rm mix} |H^{(0)}|^2 (\tilde{\varphi}^{(0)})^2
+ \cdots,
\label{L4D}
\end{eqnarray}
where we use the Fourier expansion (\ref{varphi-exp1}) for $H$
and (\ref{tildevarphi-exp}) for $\tilde{\varphi}$.

As seen from (\ref{L4D}), $C_5^{(0)}$ is massless at the tree level.
After receiving radiative corrections,
the effective potential relating to $C_5^{(0)}$ is induced
and $C_5^{(0)}$ acquires a mass through
the Hosotani mechanism~\cite{H1,H2}.
Concretely, the one-loop effective potential for the Wilson line phase 
$\theta (= \sqrt{2\pi R} \tilde{g}_5 C_5^{(0)})$
is derived as
\begin{eqnarray}
\hspace{-1.5cm}&~& V_{\rm eff}[\theta]
= -\frac{1}{2} \int \frac{d^4p_{\rm E}}{(2\pi)^4}
\sum_{n=-\infty}^{\infty} \ln\left\{
p_{\rm E}^2 + m_{\tilde{\varphi}}^2
+ \left(\frac{2\pi n + \beta_{\tilde{\varphi}} -
\tilde{q}_{\tilde{\varphi}} \theta}{2\pi R}\right)^2
\right\}
\nonumber \\
\hspace{-1.5cm}&~& ~~~~~~~~ = E_0 - \frac{3}{64\pi^6 R^4}
\sum_{n=1}^{\infty} \left(\frac{1}{n^5}
+ \frac{r_{\tilde{\varphi}}}{n^4} + \frac{r_{\tilde{\varphi}}^2}{3 n^3}\right)
e^{-n r_{\tilde{\varphi}}} \cos\left\{n \left(\beta_{\tilde{\varphi}} -
\tilde{q}_{\tilde{\varphi}} \theta\right) \right\},
\label{Veff}
\end{eqnarray}
where $p_{\rm E}$ is a 4D Euclidean momentum,
$E_0$ is a $\theta$-independent constant
and $r_{\tilde{\varphi}} = 2\pi R m_{\tilde{\varphi}}$.
The physical vacuum is realized at 
$\beta_{\tilde{\varphi}} - \tilde{q}_{\tilde{\varphi}} \theta = 0$
and $C_5^{(0)}$ decouples in the low-energy theory,
if $R$ is small enough,
by acquiring the mass of $O(1/R)$.

The scalar field $\tilde{\varphi}^{(0)}(x)$ survives
in a post-SM at the terascale for 
$\beta_{\tilde{\varphi}} - \tilde{q}_{\tilde{\varphi}} \theta = 0$
and $m_{\tilde{\varphi}} < O(1)$TeV,
and we find that our Lagrangian density agrees with 
that containing a dark matter in a specific model
called the New Minimal Standard Model (NMSM)~\cite{BPV,DKLM}.
Then, $\tilde{\varphi}^{(0)}(x)$ becomes 
a possible candidate of dark matter.

The $\tilde{\varphi}^{(0)}(x)$ couples to the SM Higgs doublet
through the quartic interaction 
$-(1/2)\lambda_{\rm mix} |H^{(0)}|^2 (\tilde{\varphi}^{(0)})^2$.
In the presence of this term as the Higgs portal,
the running of $\lambda$ based on the renormalization group equation
changes compared with that in the SM, and
the vacuum stability of Higgs potential can be improved~\cite{DKLM,HKT}.

Here, as a complementary comment on our hypothesis,
we state a feature that
{\it matters are not necessarily classified
into the visible ones and the hidden ones, 
even if a system has two $U(1)$ gauge bosons
$B_M$ and $C_M$ with different types of BCs,
because there can exist particles that possess both $U(1)$ charges.}
Let us show it using a model described by the Lagrangian density,
\begin{eqnarray}
\mathscr{L}_{\tilde{\varphi}_a}
= \sum_{a=1, 2} \left\{(D_M \tilde{\varphi}_a)^{*} (D^M \tilde{\varphi}_a) 
 - m_{\tilde{\varphi}_a}^2 |\tilde{\varphi}_a|^2\right\}
 - \frac{1}{4} B_{MN}B^{MN}
 - \frac{1}{4} C_{MN}C^{MN},
\label{La}
\end{eqnarray}
where $D_M=\partial_M-ig_5q_{\tilde{\varphi}_a} B_M 
- i\tilde{g}_5 \tilde{q}_{\tilde{\varphi}_a} C_M$ 
for a pair of complex scalar fields 
$\tilde{\varphi}_a = \tilde{\varphi}_a(x, y)$ $(a=1,2)$.
In case that $q_{\tilde{\varphi}_1} = q_{\tilde{\varphi}_2}$, $\tilde{q}_{\tilde{\varphi}_1} = -\tilde{q}_{\tilde{\varphi}_2}$
and $m_{\tilde{\varphi}_1} = m_{\tilde{\varphi}_2}$,
$\mathscr{L}_{\tilde{\varphi}_a}$ is a single-valued function
under the BCs (\ref{BM-OBC-T}),  (\ref{BM-OBC-Z2}),
(\ref{CM-CBC-T}),  (\ref{CM-CBC-Z2}) and
\begin{eqnarray}
\tilde{\varphi}_a(x, y+2\pi R)
=e^{i \beta_{\tilde{\varphi}}} \tilde{\varphi}_a(x, y),~~
\tilde{\varphi}_1(x, -y)= \eta_{\tilde{\varphi}} \tilde{\varphi}_2(x, y),
\label{tildevarphi-a-BC}
\end{eqnarray}
where $\beta_{\tilde{\varphi}}$ takes 0 or $\pi$
and $\eta_{\tilde{\varphi}}$ takes $1$ or $-1$.
We refer to the $U(1)$ gauge symmetry 
concerning the BCs (\ref{CM-CBC-T}),  (\ref{CM-CBC-Z2}) and
(\ref{tildevarphi-a-BC})
as an {\it exotic $U(1)$ symmetry}~\cite{KM1,KM2}.\footnote{
The orbifolding due to these BCs is regarded as 
a variant of the diagonal embedding proposed in
\cite{KTY}.
}
Then, we find that 
$\tilde{\varphi}_a$ own both $U(1)$ gauge quantum numbers.
A similar feature holds on a theory containing non-abelian gauge symmetries:
matters can possess both gauge quantum numbers
whose gauge bosons satisfy different types of BCs
if the theory is vector-like.

\section{Gauge-higgs inflation}
\label{GHI}

We apply a model with conjugate BCs on a gauge-Higgs inflation scenario.
Let us consider a gravity theory 
coupled to a $U(1)_C$ gauge theory 
defined on a 5D space-time whose classical background 
is $M^4 \times S^1/Z_2$.
The starting action is given by
\begin{eqnarray}
\hspace{-1cm}&~& S_{\rm 5D}^{\rm gr}
= \int d^5x\sqrt{-\hat{g}_5}\Bigg[\frac{1}{16\pi G_5}\hat{R_5}
-\frac{1}{4}\hat{g}^{MP}\hat{g}^{NL} C_{MN} C_{PL} 
\nonumber \\
\hspace{-1cm}&~& ~~~~ \left.
+ \sum_{a=1}^{c_1} \bar{\tilde{\psi}}_a^{\rm n}
(-i\hat{g}^{MN} \hat{\Gamma}_M \nabla_N-\mu_a) \tilde{\psi}_a^{\rm n}
+ \sum_{b=1}^{c_2}\bar{\tilde{\psi}}_b^{\rm ch}
(-i\hat{g}^{MN} \hat{\Gamma}_M D_N-m_b) \tilde{\psi}_b^{\rm ch}
\right],
\label{S-5D-gr}
\end{eqnarray}
where $\hat{g}_5=\det\hat{g}_{MN}$, 
$\hat{g}^{MN}$ is the inverse of 5D metric $\hat{g}_{MN}$, 
$G_5$ is the 5D Newton constant,
$\hat{R}_5$ is the 5D Ricci scalar,
$C_{MN}=\partial_M C_N-\partial_N C_M$, 
$\hat{\Gamma}_M = E_M^k \Gamma_k$
($E_M^k = E_M^k(x, y)$ is the f$\ddot{\rm u}$nf bein, 
$\Gamma_k$ are 5D gamma matrices,
and $k$ is the space-time index in the local Lorentz frame),
$\nabla_N=\partial_N - (i/4) \hat{\omega}_{N}^{kl}\Sigma_{kl}$
($\hat{\omega}_N^{kl}$ is the spin connection 
and $\Sigma_{kl} =i[\Gamma_k, \Gamma_l]/2$),
$D_N=\partial_N - (i/4) \hat{\omega}_{N}^{kl}\Sigma_{kl}
-i \tilde{g}_5 \tilde{q}_b C_N$ for $\tilde{\psi}_b^{\rm ch}$, 
$C_N$ is a 5D $U(1)_C$ gauge boson in the hidden sector
and we assume that it satisfies the conjugate BCs 
(\ref{CM-CBC-T}) and (\ref{CM-CBC-Z2}),
$\tilde{\psi}_a^{\rm n}$ are neutral fermions, 
$\tilde{\psi}_b^{\rm ch}$ are $U(1)_C$ charged fermions
whose $U(1)_C$ charge is $\tilde{q}_b$,
and $c_1$ and $c_2$ stand for numbers of
neutral and charged fermions, respectively.
The $\tilde{g}_5$ is a 5D gauge coupling constant.

If the SM gauge bosons satisfy the ordinary BCs
such as (\ref{BM-OBC-T}) and (\ref{BM-OBC-Z2})
and both $\tilde{\psi}_a^{\rm n}$ and $\tilde{\psi}_b^{\rm ch}$
satisfy the BCs (\ref{psi-CBC}) with 
$\beta_a$ and $\beta_b$ as a twisted phase ($\beta_{\tilde{\psi}}$),
$\tilde{\psi}_a^{\rm n}$ and $\tilde{\psi}_b^{\rm ch}$
should be singlets of the SM gauge group,
as a consequence in the previous section.

The BCs of $\hat{g}_{MN}$ are given by
\begin{eqnarray}
\hspace{-1cm} &~& \hat{g}_{MN}(x, y+2\pi R)=\hat{g}_{MN}(x, y),
\label{BC-gr-T}\\
\hspace{-1cm} &~& \hat{g}_{\mu\nu}(x, -y)=\hat{g}_{\mu\nu}(x, y),~~ 
\hat{g}_{\mu5}(x, -y)=-\hat{g}_{\mu5}(x, y),~~ 
\hat{g}_{55}(x, -y)=\hat{g}_{55}(x, y),
\label{BC-gr-Z2}
\end{eqnarray}
and then the Fourier expansions of $\hat{g}_{MN}$ are presented as
\begin{eqnarray}
&~& \hat{g}_{\mu\nu}(x, y)
=\hat{g}_{\mu\nu}^{(0)}(x) +  
\sum_{n=1}^{\infty}\hat{g}_{\mu\nu}^{(n)}(x)\cos\frac{ny}{R}, 
\label{gmunu-exp}\\
&~& \hat{g}_{\mu5}(x,y)
=  \sum_{n=1}^{\infty}\hat{g}_{\mu5}^{(n)}(x)\sin\frac{ny}{R}y, 
\label{gmu5-exp}\\
&~& \hat{g}_{55}(x,y)
= \hat{g}_{55}^{(0)}(x) +
 \sum_{n=1}^{\infty}\hat{g}_{55}^{(n)}(x)\cos\frac{ny}{R}.
\label{g55-exp}
\end{eqnarray}

The spin connection $\hat{\omega}_M^{kl}$ satisfy 
the ordinary BCs such that
\begin{eqnarray}
&~& \hat{\omega}_M^{kl}(x, y+2\pi R)=\hat{\omega}_M^{kl}(x, y),
\label{BC-spin-T}\\
&~& \hat{\omega}_\mu^{kl}(x, -y)=\hat{\omega}_\mu^{kl}(x, y),~~ 
\hat{\omega}_5^{kl}(x, -y)=-\hat{\omega}_5^{kl}(x, y),
\label{BC-spin-Z2}
\end{eqnarray}
and then the full Lagrangian density containing both visible
and hidden sectors becomes a single-valued function
on $S^1/Z_2$.

On the Minkowski background, $\hat{g}_{\mu\nu}^{(0)}$
takes the classical value such as 
$\langle g_{\mu\nu}^{(0)}\rangle=\eta_{\mu\nu}$,
and other zero modes are assumed to have the following classical values: 
\begin{align}
\langle \hat{g}_{55}^{(0)} \rangle=\phi^{2/3},~ 
\langle C_{5}^{(0)} \rangle=\frac{\theta}{\sqrt{2\pi R} \tilde{g}_5},
\end{align}
where $\phi$ is the radion and $\theta$ is the Wilson line phase.
The Kaluza-Klein modes are assumed to have zero classical values.

According to a usual procedure,
the following effective potential is obtained at the one-loop level,
\begin{eqnarray}
\hspace{-1.5cm}&~& V(\rho,\theta)=\frac{3L^{2}m^{6}}{2 \pi^{2}\rho^{2}}\left[-2\zeta(5)+c_{1}\sum^{\infty}_{n=1}\left(\frac{1}{n^{5}}+r_{m}\frac{\rho^{1/3}}{n^{4}}+r^{2}_{m}\frac{\rho^{2/3}}{3n^{3}}\right)
e^{-nr_{m}\rho^{1/3}}\right.
\nonumber\\
\hspace{-1.5cm}&~&~~~~~~\left.+c_{2}\sum^{\infty}_{n=1}\left(\frac{1}{n^{5}}+\frac{\rho^{1/3}}{n^{4}}+\frac{\rho^{2/3}}{3n^{3}}\right)e^{-n\rho^{1/3}}
\cos\left\{n\left(\beta - \tilde{q}\theta\right)\right\}\right]
+\frac{L^{2}m}{\rho^{1/3}} \tilde{a} + \cdots,
\label{V-inf}
\end{eqnarray}
where we take common masses $\mu = \mu_a$ and
$m = m_b$, a common twisted phase $\beta = \beta_b$
and a common charge $\tilde{q} = \tilde{q}_b$
for simplicity, $L = 2\pi R$, $\rho= L^3 m^3 \phi$,
$\zeta(k)=\sum_{n=1}^{\infty}1/{n^k}$,
$r_m = \mu/m$ and $\tilde{a}$ is some constant.

The above potential has a same form as that obtained in \cite{AIKK1}
except overall factor and $\beta$, and hence
both radion and Wilson line phase are stabilized in case with $c_1 > 2 + c_2$,
and $\theta$ is, in particular, fixed as
$\beta - \tilde{q} \theta = \pi$.
Furthermore, the gauge-Higgs field $\theta$ can give rise to inflation
in accord with the astrophysical data~\cite{AIKK2}.

We need some modification of our model
to explain the origin of the Big Bang 
after inflaton decays into the SM particles.
The direct coupling 
between inflaton and some SM particles
is necessary to produce radiations at a very early universe, 
but it is difficult due to the mismatch of BCs,
as explained in the previous section.
As a way out, if some SM particles or its extension form 
a pair of vector-like multiplet for $U(1)_C$
and satisfy the BCs such as (\ref{tildevarphi-a-BC})
or counterparts of fermions,
they can directly couple to $C_5^{(0)}$.
For instance, if there exist two Higgs doublets $H_a$  
as a vector-like pair of $U(1)_C$,
there can appear the coupling such as 
$\tilde{g}_5 ^2 \tilde{q}_H^2 |H_a^{(0)}|^2 (C_5^{(0)})^2$.
In this case, 
although the contributions from $H_a$ are added
to the potential (\ref{V-inf}), $\theta$ might remain
inflaton because they are not dominated.

\section{Conclusions and discussions}
\label{CD}

We have formulated 5D $U(1)$ gauge theories
yielding conjugate BCs besides ordinary ones on $S^1/Z_2$.
On the conjugate BCs, the 4D components of $U(1)_C$ gauge boson
have odd $Z_2$ parities and their zero modes are projected out
through the dimensional reduction.
Then, the $U(1)_C$ gauge symmetry is broken down by orbifolding.
In contrast, the fifth component of $U(1)_C$ gauge boson
has even $Z_2$ parities, and its zero mode $C_5^{(0)}$ survives
and becomes a dynamical field.
It is massless at the tree level, 
but after receiving radiative corrections,
the effective potential relating to $C_5^{(0)}$ is induced
and $C_5^{(0)}$ acquires a mass of $O(1/R)$
($R$ is the radius of $S^1$)
and decouple to the low-energy theory
if $R$ is small enough.
Matter fields transform into the charge conjugated
ones under the $Z_2$ transformation, 
in case yielding the conjugate BCs.
Then, only real fields such as real scalar fields
and Majorana fermions appear after compactification.

We have shown that the separation of visible and hidden particles
can be realized in gauge interactions 
using a 5D extension of the SM with an extra $U(1)$
gauge symmetry and an extra scalar field coexisting different types of BCs,
and derived the Lagrangian density containing 
a dark matter in the NMSM.
The zero mode of extra scalar field yielding the conjugate BCs
becomes a possible candidate of dark matter.

Furthermore, we have applied a 5D gravity theory
coupled to a $U(1)$ gauge theory with conjugate BCs 
on a gauge-Higgs inflation scenario and
found that the effective potential containing the radion $\phi$ 
and Wilson line phase $\theta$
plays a role of an inflaton potential 
and $\theta$ becomes inflaton.

We have studied a model of dark matter 
and that of inflaton independently, 
to demonstrate the origin of such hidden particles clearly.
We can combine them into
a 5D gravity theory coupled to an extension of
the SM (or grand unified theory) containing 
an extra $U(1)$ gauge boson $C_M$, 
an extra scalar field $\tilde{\varphi}$ and some extra fermions.
In the combined theory, the zero mode of $\tilde{\varphi}$ 
becomes dark matter with 
$\beta_{\tilde{\varphi}} - \tilde{q}_{\tilde{\varphi}} \theta = 0$,
and that of $C_5$ becomes inflaton
with $\beta - \tilde{q} \theta = \pi$.
Then, $\theta$ must satisfy 
$\theta = \beta_{\tilde{\varphi}}/\tilde{q}_{\tilde{\varphi}}
= (\beta - \pi)/\tilde{q}$, and hence the problem of origin of hidden particles
is replaced by that of the selection for twisted phases
($\beta_{\tilde{\varphi}}, \beta$) 
and $U(1)$ charges ($\tilde{q}_{\tilde{\varphi}}, \tilde{q}$)
to fix a suitable value of $\theta$.
It would be interesting to construct a more realistic model
based on the combined one.
To generate the state of Big Bang after the end of inflation,
it is necessary to couple some SM particles to inflaton
and it can be realized,
if some SM particles have an exotic $U(1)$ charge
and they are related to ones with an opposite charge
under the $Z_2$ reflection.

Finally, we give a comment on the right-handed neutrinos.
Because the right-handed neutrinos are singlets of the SM gauge group
and they have Majorana masses,
we guess that they might be hidden matters obeying conjugate BCs.
But it is difficult to realize it,
because we cannot construct a $Z_2$ invariant term 
in 5D Lagrangian density 
to derive the 4D Yukawa interaction relating to neutrinos,
due to the mismatch of BCs between the SM non-singlets and singlets.
Nevertheless, it would also be interesting to examine the origin of 
the right-handed neutrinos from the viewpoint of BCs.

\section*{Acknowledgments}
Y.~A. wishes to thank Iwanami Fu-Jukai for his Research grant
and he would like to thank Prof. Katsushi Ito for his kind hospitality.


\begin{thebibliography}{99}
\bibitem{HM}
A.~Hebecker and J.~March-Russell, Nucl. Phys. B {\bf 625}, 128 (2002).

\bibitem{HKO} 
N.~Haba, Y.~Kawamura and K.~Oda, Phys. Rev. D {\bf 78}, 085021 (2008).

\bibitem{AGGJ}
A.~Ahmed, B.~Grzadkowski, J.~F.~Gunion and Y.~Jiang,
JHEP {\bf 1510}, 033 (2015).

\bibitem{H1}
Y.~Hosotani, Phys.~Lett. B {\bf 126}, 309 (1983).

\bibitem{H2}
Y.~Hosotani, Ann. of Phys. {\bf 190}, 233 (1989).

\bibitem{BPV}
C.~P.~Burgess, M.~Pospetov and T.~ter~Veldhuis, 
Nucl. Phys. B {\bf 619}, 709 (2001).

\bibitem{DKLM}
H.~Davoudiasl, R.~Kitano, T.~Li and H.~Murayama,
Phys.~Lett. B {\bf 609}, 117 (2005).

\bibitem{HKT}
N.~Haba, K~Kanata and R.~Takahashi,
JHEP {\bf 1404}, 029 (2014).

\bibitem{KM1} 
Y.~Kawamura and T.~Miura, Int. J. Mod. Phys. A {\bf 26}, 4405 (2011).

\bibitem{KM2} 
Y.~Kawamura and T.~Miura, Int. J. Mod. Phys. A {\bf 27}, 1250023 (2012).

\bibitem{KTY}
K.~Kojima, K.~Takenaga and T.~Yamashita, 
Phys. Rev. D {\bf 84}, 051701(R) (2011).

\bibitem{AIKK1} 
Y.~Abe, T.~Inami, Y.~Kawamura and Y.~Koyama,
Prog. Theor. Exp. Phys. {\bf 2014}, 073B04 (2014).

\bibitem{AIKK2} 
Y.~Abe, T.~Inami, Y.~Kawamura and Y.~Koyama,
Prog. Theor. Exp. Phys. {\bf 2015}, 093B03 (2015).

\end{thebibliography}
\end{document}